\documentclass[aps]{revtex4}
\usepackage{amssymb}
\usepackage{graphicx}
\usepackage{amsmath}
\begin{document}
\title{Cauchy  flights in confining potentials}
\author{Piotr Garbaczewski}\email{pgar@uni.opole.pl}
\affiliation{
Institute of Physics, University of Opole, 45-052 Opole, Poland}
\begin{abstract}
   We   analyze   confining  mechanisms  for  L\'{e}vy  flights evolving under  an  influence of  external  potentials.
    Given a stationary   probability density function (pdf), we address the  reverse engineering problem:  design a jump-type stochastic  process
    whose target  pdf (eventually asymptotic) equals the preselected one.
 To this end, dynamically distinct jump-type processes can be employed.  We demonstrate that one "targeted stochasticity"
   scenario involves  Langevin  systems with a symmetric stable  noise. Another derives from   the  L\'{e}vy-Schr\"{o}dinger
    semigroup dynamics  (closely linked with topologically induced super-diffusions), which has  no standard   Langevin  representation.
  For computational and visualization purposes, the Cauchy driver is employed to   exemplify our considerations.
    \end{abstract}

 \maketitle
\vskip0.2cm
\section{Motivation}

We consider a subclass of so-called L\'{e}vy flights that is   mathematically  identifiable  as symmetric stable
stochastic processes, \cite{levy}-\cite{sato}. These non-Gaussian processes are of the jump-type and, in contrast  to familiar
Gaussian diffusion-type processes, involve a number of  obstacles.

 One of them is a shortage of explicit analytic solutions. e.g.
explicit  probability density functions (pdfs) and transition probability densities.
Another comes from theoretically admitted existence of arbitrarily small jumps
 and fat tails of the pdf which  typically  preclude the existence of (in the least the  second)  moments. In the presence of
  confining potentials the resultant pdfs may admit higher  moments, but merely a finite number of them  may exist.
  Third, some care is needed  in any computer-assisted analysis of L\'{e}vy flights, since imposing  a lower and upper bound on the size
  of jumps,  sets the problem within the ramifications of the  central limit theorem which   implies the standard Gaussian limit,
   \cite{stanley}, albeit the simulated process is non-Gaussian by definition.

In the present paper we  set  general confinement criterions for  symmetric stable processes, with special emphasis  on the  analytically
tractable case of the Cauchy noise \cite{fogedby}-\cite{dubkov2}. With regard to a specific  response to external potentials, we pay  special  attention to two  classes of  jump-type processes: those  related to  the  Langevin
equation  and those  induced  by the Cauchy-Schr\"{o}dinger  semigroup dynamics (that involves  a fractional analog of the generalized
diffusion equation). We leave for a separate study another interesting possibility, that is based on Bochner's concept of subordination,
c.f. \cite{eliazar}.

L\'{e}vy  flights in confining potentials  with a standard (mostly for an additive noise)  Langevin representation have
 received ample  attention, \cite{fogedby}-\cite{cufaro} and \cite{dubkov,dubkov2}.
 For   Cauchy-Langevin processes, a manipulation with  the forward drift and/or
 its (external conservative force) potential sets rules of the game, e.g.  directly  leads to stationary probability densities.
 They   never have a Gibbs-Boltzmann form, characteristic for Gaussian diffusion processes, \cite{eliazar}.

 Another class of  L\'{e}vy  processes, that are driven by dynamical semigroups,    was analyzed in  detail in \cite{klauder,olk}. 
  Cauchy driver has received  there special attention.  The semigroup-driven  processes  independently reappeared (in the context   of systems with topological complexity like folded polymers or complex networks)   in Refs. \cite{brockmann}-\cite{geisel1}.  
In Refs.  \cite{klauder,olk},  external potentials appear as additive perturbations of the noise generator and under the name of
 effective potentials  they appear in  \cite{brockmann} as well.  In the "topological" literature, the resultant semigroup dynamics has been
  implemented via local modifications of jump rates of the associated jump-type  process. 
  
  The semigroup-driven (and topologically-induced)  processes  appear to have  no
  standard  Langevin representation, irrespective of whether we adopt  an  additive or multiplicative noise. Only under special circumstances a connection with the   multiplicative noise has been established in Ref. \cite{brockmann}, but it is not a generic property of semigroup-driven processes, see e.g. also \cite{srokowski,srokowski1}.

 Although we formulate a framework incorporating symmetric stable processes in general, we strongly rely on a  mathematical  theory  of the
  Cauchy  semigroup-driven processes. This theory, without any   "topological" context,   has been formulated in Ref. \cite{olk}.
  No explicit  analytic examples  of confining potentials, nor pdfs  were given there.

We  shall demonstrate that  the Langevin-driven  and   semigroup-driven  L\'{e}vy   processes  stay in  affinity and may share common  for
both stationary pdf.  A super-diffusive   dynamical pattern of behavior is  generically expected to arise. An  asymptotic approach
  towards a stationary pdf is  then   in principle possible.

  This motivates our  "targeted stochasticity"  discussion  whose   original formulation for Langevin-driven  L\'{e}vy  systems can
   be found in Ref. \cite{eliazar},  the   reverse engineering problem  being included.  The necessity of considering the semigroup dynamics in this
   reconstruction problem points first, to an independent dynamical mechanism,  and second, to the fact that only the  properly tailored semigroup
    (and thus the semigroup potential) may guarantee that the prescribed invariant density actually \it is \rm an asymptotic target of the process.
     This point was left aside in "topological" references \cite{brockmann}-\cite{geisel1}, where the pertinent invariant densities were postulated
     to have a Gibbsian form, hence a tacit assumption was made about extyremely strong confining properties of the topologically-induced process.

The original  reverse engineering  problem reads:  given a  stationary pdf, can we tailor a drift function so that
the system Langevin dynamics would  admit the predefined as an asymptotic target ?
 In the course of our  discussion, we in fact  extend the  range of applicability
 of the original "targeted stochasticity"  scenario  and   demonstrate  that, for a symmetric stable driver,
   a priori chosen stationary pdf  may serve as  a target density for   {\it  both}  Langevin and semigroup-driven  jump-type processes.
   Even   though their  detailed  dynamical patterns of behavior are different.
   In the near-equilibrium regime this  dynamical distinction  becomes immaterial.
     For  analytically tractable  and  visualization  insights, we shall  basically refer to
 Cauchy processes   in  confining  potentials.

For the record we point out that the term "equilibrium" needs to be addressed with some care for non-Gaussian processes.
No physical thermalization  mechanisms  have ever  been proposed  for
 L\'{e}vy  flights. Moreover, their physical "reason" (origin of noise)  appears to be  exterior  to the physical system,
  with no  reliable kinetic  theory background,   and therefore  no   fluctuation-dissipation response theory
  could have been   been set for  any stable noise.

  To the contrary, the noise  "reason"  is definitely an intrinsic feature of the
 environment-particle coupling in case of the standard  Brownian motion, based on the kinetic theory derivations.
 All traditional fluctuation-dissipation  relationships  find their place in the Brownian framework. None of them has been  reproduced in the context of L\'{e}vy flights.

\section{Response of Gaussian  noise   to confining  potentials: Smoluchowski processes}

Albeit we are primarily interested in  jump-type stochastic processes, certain useful  intuitions can be borrowed
  from the standard theory of Brownian motion.  Namely, let us consider a one-dimensional Smoluchowski
diffusion process \cite{risken}, with the Langevin representation
\begin{equation}
\dot{x} = b(x,t) +A(t)
\end{equation}
 where $\langle A(s)\rangle =0$, $\langle
A(s)A(s')\rangle = {2D} \delta (s-s')$ and  $b(x)$ is a forward
drift of the process having the gradient form $b= 2D \nabla \Phi $,
where $D$ stands for a diffusion constant.

If an initial probability density $\rho_0(x)$ is given, then the
diffusion process  obeys  the Fokker-Planck
equation
\begin{equation}
 \partial _t\rho = D\Delta \rho - \nabla \, ( b \cdot
\rho )\, .
\end{equation}

We introduce an osmotic  velocity  field  $u = D\ln \rho $,
together with the current velocity field  $v=b - u$. The latter
obeys  the continuity equation $\partial _t \rho = - \nabla j$,
where  $j= v\cdot \rho $  has   a standard interpretation of a
probability current.

Presently we   pass to time-independent drifts of the  diffusion process,
 that are induced by   external (conservative,  Newtonian) force fields $f= - \nabla V$.
One arrives at Smoluchowski diffusion processes by  setting
 \begin{equation}
b = {\frac{f}{m\beta }} = - {\frac{1}{m\beta }}  \nabla V \, .
\end{equation}
This  expression accounts for the fully-fledged phase-space derivation of the spatial process, in the regime of large $\beta $.
 It is  taken  for granted that  the fluctuation-dissipation
  balance gives rise to the  standard form $D=k_BT/m\beta $ of the  diffusion coefficient.

Let us consider a stationary asymptotic  regime, where $j\rightarrow j_*=0$.
We denote an  (a priori assumed to exist) invariant density    $\rho _*= \rho _*(x) $.
Since   $v_*=0$ and  $b=f/m\beta  $,  by its very definition,  does  not   depend functionally on the probability density, there holds
\begin{equation}
b=b_*=u_* = D \nabla  \ln \rho _*  \, .
\end{equation}
Consequently, we have  $  \rho _*(x) = (1/Z)\,  \exp[ - V(x)/k_BT]$, where $1/Z$ is a normalization constant.
Our outcome   has   the  familiar   Gibbs-Boltzmann  form.

Following a standard procedure \cite{risken,gar}  we transform the
Fokker-Planck equation into an associated Hermitian (Schr\"{o}dinger-type)  problem by means of a  redefinition
\begin{equation}
\rho (x,t) = \Psi (x,t) \rho _*^{1/2}(x)
\end{equation}
 that takes the Fokker-Plack equation into  a parabolic one, often called a generalized diffusion   equation:
\begin{equation}
 \partial
_t\Psi= D \Delta \Psi - {\cal{V}} \Psi \, .
\end{equation}
 Its potential ${\cal{V}}$  derives, as a function of the drift $b(x)$,   from a compatibility condition
\begin{equation}
{\cal{V}}(x) = (1/2)[b^2/(2D) + \nabla b] \, .
\end{equation}
In view of Eq. (4), its equivalent form is
\begin{equation}
{\cal{V}}(x) =  D {\frac{\Delta \rho _*^{1/2}}{\rho _*^{1/2}}}.  \label{compatibility}
\end{equation}

If the  ($1/2mD$ rescaled) Schr\"{o}dinger-type  Hamiltonian
$\hat{H} = -D\Delta + {\cal{V}}$ is a  bounded from below,
self-adjoint operator in a suitable Hilbert space, then one arrives
at a  dynamical semigroup  $\exp(-t\hat{H})$,  with the dynamical
rule $ \Psi(x,t) = [ \exp(-t\hat{H})\Psi](x,0)$, pushing
forward  in time  the   initial data $\Psi(x,0)$. The semigroup is
contractive, hence asymptotically $\Psi (x,t) \rightarrow \rho
_*^{1/2}$. Accordingly,  $\rho (x,t)\rightarrow \rho _*(x)$.

The  above Schr\"{o}dinger semigroup (parabolic) reformulation of the Fokker-Planck equation  refers to
the very same  diffusion process and the dynamics of $\rho(x,t)$ does not depend on the theoretical framework of choice.
 In below we shall demonstrate that for non-Gaussian processes,  the  semigroup-driven  and Langevin-induced dynamics
  refer to inequivalent dynamical patterns of behavior. Even if  both  are associated with a common  stationary (target) pdf.

\section{Response of L\'{e}vy flights  to confining potentials}

\subsection{L\'{e}vy   driver}

Let us set
general rules of the game with respect to the  response of any symmetric
 stable noise to external potentials. We recall that a
characteristic function of a random variable $X$  completely
determines a probability distribution of that variable. If this
distribution admits a density $\rho(x)$, we can write $<\exp(ipX)> =
\int_R \rho (x) \exp(ipx) dx$ which, for infinitely divisible
probability laws,  gives rise to  the famous L\'{e}vy-Khintchine
formula (see, e.g. \cite{applebaum}).

From now  on, we concentrate on the integral part of the
L\'{e}vy-Khintchine formula, which is responsible for arbitrary
stochastic jump features:
\begin{equation}
F(p) = - \int_{-\infty }^{+\infty } \left[\exp(ipy) - 1 -
\frac{ipy}{ {1+y^2}}\right] \nu (dy),\label{g3}
\end{equation}
where $\nu (dy)$ stands for the appropriate L\'{e}vy measure. The
corresponding non-Gaussian Markov process is characterized by
$<\exp(ipX_t)>= \exp[-t F(p)]$ and, upon setting  $\hat{p} = - i\nabla $
 instead of $p$,  yields an operator $F(\hat{p})=\hat{H}$ which is a direct analog of the free Schr\"{o}dindger Hamiltonian.

We restrict further  considerations to non-Gaussian random variables
whose probability densities are centered and symmetric, e.g.  a
subclass of stable distributions characterized by
\begin{equation}
F(p) = \lambda   |p|^{\mu } \Rightarrow \hat{H} \doteq   \lambda
|\Delta |^{\mu /2}.\label{g4}
\end{equation}
(In passing, we note that the adopted  definition of a  pseudo-differential operator
may be replaced by the negative of  a suitable Riesz fractional derivative.)
In the above,  $\mu <2$ and $\lambda >0$ stands for the intensity parameter
of the L\'{e}vy  process. The  fractional Hamiltonian   $\hat{H}$,
which is a non-local pseudo-differential operator, by construction
is positive and self-adjoint on a properly tailored domain.  A
sufficient and necessary condition for both these  properties to
hold true is that the pdf of the  L\'{e}vy process is symmetric,
\cite{applebaum}.

The associated  jump-type   dynamics is interpreted in terms of
L\'{e}vy flights. In particular
\begin{equation}\label{g5}
F(p)= \lambda  |p| \to \hat{H}= F(\hat{p}) =  \lambda |\nabla |
\doteq \lambda (-\Delta )^{1/2}
\end{equation}
refers to the Cauchy process, see e.g. \cite{klauder,olk,olk1}.

The pseudo-differential Fokker-Planck equation, which  corresponds
to the fractional Hamiltonian  \eqref{g4} and the fractional
semigroup $\exp(-t\hat{H}_{\mu })=\exp(-\lambda |\Delta |^{\mu
/2})$, reads
\begin{equation}\label{g6}
\partial _t \rho  = -  \lambda |\Delta |^{\mu /2} \rho  \, ,
\end{equation}
to be compared with the Fokker-Planck equation for freely diffusing
particle  $\partial _t \rho = D \Delta
\rho $.

For a pseudo-differential operator $|\Delta |^{\mu /2}$, the action
on a function from  its domain is  greatly simplified, in view of
the properties of the  L\'{e}vy measure $\nu _{\mu }(dx)$.  Namely, remembering that we overcome a
 singularity at $0$  by means of the  principal value  of the  integral, we have
\cite{klauder,olk}:
\begin{equation}
 (|\Delta |^{\mu /2} f)(x)\, =\, - \int  [f(x+y) - f(x) ] \nu _{\mu }(dy) \, .
\end{equation}
By changing  an integration variable $y$  to $z=x+y$ and employing
 a direct connection  with the Riesz fractional derivative of the $\mu $-th order, \cite{dubkov},
  we  arrive at
\begin{equation}
 ( |\Delta |^{\mu /2} f)(x)\, =\, -  {\frac{\Gamma (\mu +1) sin(\pi \mu/2)}{\pi }} \int  {\frac{f(z)- f(x)}{|z-x|^{1+\mu }}}\,
 dz \,
\end{equation}
with $( |\Delta |^{\mu /2} f)(x) =  -  \partial ^{\mu }f(x)/\partial |x|^{\mu }$.
The case of $\mu =1$  refers to the  Cauchy driver (e.g. noise).

 We note   a systematic sign difference between
our notation for a pseudo-differential operator $|\Delta |^{\mu /2}$ and this based on the fractional derivative notion,
like e.g.  $ \Delta ^{\mu /2} \doteq     \partial ^{\mu }/\partial |x|^{\mu }$ of Refs. \cite{brockmann,geisel}.

\subsection{Langevin scenario}

In case of jump-type (L\'{e}vy) processes a response to external
perturbations by conservative force fields appears to be
particularly interesting. On the  one hand, one encounters  a widely
accepted reasoning (Refs. \cite{fogedby}-\cite{dubkov}) where the
Langevin equation, with additive deterministic and  L\'{e}vy "white
noise" terms,  is found to imply a fractional Fokker-Planck
equation, whose form  faithfully  parallels  the Brownian  version,
e.g. (c.f. \cite{fogedby}, see also \cite{olk1})
$$
\dot{x}= b(x)  + A^{\mu }(t)
$$
$$
\Downarrow
$$
\begin{equation}\label{g7}
 \partial _t\rho = -\nabla (b\cdot \rho ) - \lambda |\Delta |^{\mu /2}\rho \, .
 \end{equation}
We emphasize a difference in sign in the
second term, if compared with Eq. (4) of Ref. \cite{fogedby}.  There, the minus sign is  absorbed in the adopted definition
of the (Riesz) fractional derivative. Apart from the formal resemblance of operator symbols, we do not directly
 employ fractional derivatives in our formalism.

Let us assume that the fractional Fokker-Planck equation \eqref{g7}
admits a stationary pdf $\rho _*(x)$. Then,  a functional form of the drift $b(x)$
can be reconstructed by means of  an indefinite integral
\begin{equation}
b(x)= - \lambda   {\frac{\int  |\Delta |^{\mu /2}\rho _*(x)\, dx}{\rho _*(x)}} \, .
\end{equation}
This is the reverse engineering problem of \cite{eliazar}.

\subsection{L\'{e}vy-Schr\"{o}dinger semigroup}

 On the other hand, by mimicking the previous Gaussian strategy,  we
can directly refer  to  the Hamiltonian framework and dynamical semigroups with
L\'{e}vy generators being additively perturbed by a suitable
potential, see e.g. \cite{klauder,olk}.
For example, assuming that the functional form  of
${\cal{V}}(x)$ guarantees that $\hat{H}_{\mu }  \doteq   \lambda
|\Delta |^{\mu /2} +  {\cal{V}}$  is self-adjoint and bounded from
below in a suitable Hilbert space, we may readily  pass to  the  fractional  (non-Gaussian, jump process) analog of the generalized diffusion equation:
\begin{equation}\label{g8}
\partial _t\Psi = -  \lambda |\Delta |^{\mu /2} \Psi
- {\cal{V}} \Psi .
\end{equation}
The dynamical semigroup reads $\exp(- t\hat{H}_{\mu })$ and the
compatibility condition affine to that of  Eq. (\ref{compatibility}),  typically  takes the form of the time-adjoint equation  for an auxiliary function $\theta (x,t)$:
\begin{equation}
\partial _t\theta = \lambda |\Delta
|^{\mu /2} \theta + {\cal{V}} \theta  \, .
\end{equation}

General theory \cite{klauder,olk}  tells us that   a   (properly normalized)   product  $\theta ^*(x,t) \theta (x,t)$   determines  a probability density  $\rho (x,t)$  of a   Markov   process that interpolates between the  boundary data $\rho(x,0)$ and $\rho (x,T)$, in the time span  $t\in [0,T]$.

Let us assume that there exists an invariant (stationary) pdf   $\rho _*$ of this Cauchy semigroup-induced process.  If we demand that $\theta (x,t)$ actually does not depend on time,  and adopt a decomposition  $\rho = \Psi \rho _*^{1/2}$, c.f.  Eq. (5),  we
are allowed to set $\theta \equiv \rho _*^{1/2}$  and remove limitations upon the  time interval.
 Ultimately,we arrive at  a compatibility condition that is a direct fractional version of Eq. (\ref{compatibility}):
\begin{equation}
{\cal{V}}  =   -\lambda\,  {\frac{|\Delta |^{\mu /2}\,  \rho ^{1/2}_*}{\rho ^{1/2}_*}} \, .
\end{equation}

In the present case, we can readily evaluate the dynamics of $\rho (x,t) = \Psi (x,t) \rho _*^{1/2}(x)$:
\begin{equation}
\partial _t \rho  = \rho _*^{1/2}\partial _t \Psi = -   \lambda   \rho _*^{1/2} |\Delta |^{\mu /2}[ \rho _*^{-1/2}
    \rho ]
    + {\cal{V}} \cdot \rho  \, .
\end{equation}
This  is a departure point for the reverse engineering procedure: given the stationary pdf, find the semigroup potential ${\cal{V}}(x)$.

It is interesting to observe that by making cosmetic changes: set $\lambda =1$, formally
identify   $\rho _*^{1/2}= \exp [- \beta  V(x)] $, with whatever $V$ and $\beta =1/k_BT$, we and up with  a familiar  form of the  transport equation  previously
  introduced  in  a number of papers:
\begin{equation}
\partial _t \rho =-    \exp(-\beta V/2)\,  |\Delta |^{\mu /2} \exp(\beta V/2 )  \rho  +
\rho \exp (\beta V/2) |\Delta |^{\mu /2} \exp(-\beta V/2) \, ,
\end{equation}
 c.f. formula (6) in \cite{geisel}, formula  (5) in \cite{geisel1} and  formula  (36) in \cite{brockmann}.
There,   the investigated process was named a topologically induced super-diffusion.  We point out  a systematic  sign  difference
between  our  $|\Delta |^{\mu /2}$   and the corresponding fractional derivative  $\Delta ^{\mu /2}$
   of \cite{brockmann,geisel,geisel1}. Graphically  these symbols look similar, but have different origin.

{\bf  Remark:} It is of some interest to  invoke an independent approach of Refs. \cite{sokolov,geisel} where
one modifies jumping rates by suitable local factors,  to arrive  at a response  mechanism that is characteristic
of the  previously outlined semigroup dynamics.   In view of (14), the  free transport equation  $\partial _t \rho  = -  \lambda |\Delta |^{\mu /2} \rho $ can be  re-written as a master equation $
\partial _t \rho (x)  = \int [w(x|z) \rho (z) - w(z|x) \rho (x)] \nu _{\mu }(dz)$. 
 The jump rate  $ w(x|y)\sim 1/|x-y|^{1+\mu }$ is  an even function,
$w(x|z)= w(x|z)$.  If we replace  the   jump rate $ w(x|y)$
     of the free fractional dynamics     by  the expression
$ w_{\phi }(x|y)\sim {\frac{\exp [\Phi (x) - \Phi (y)]}{|x-y|^{1+\mu }}}$ and
account for the fact that $w_{\phi }(x|z) \neq w_{\phi }(z|x)$,  then  the master equation  takes the form:
$(1/\lambda )\partial _t \rho = |\Delta |^{\mu /2}_{\Phi }f =
-  (\exp \Phi )\, |\Delta |^{\mu /2}[ \exp(-\Phi )
    \rho ]   +
     \rho \exp (-\Phi ) |\Delta |^{\mu /2} \exp(\Phi ) $.
Whatever $\Phi (x) $ has been chosen (up to a normalization factor), then  formally
 $\rho _* =\exp 2\Phi $ is a stationary solution of  that transport eqution.
We note that a  physically attractive point in the topologically-induced dynamics pattern was an assumption that
$\exp 2\Phi $ sets a Gibbsian form of the pdf. Accounting for the normalization factor $1/Z$ one
presumes that   $\rho _* =(1/Z) exp(-V_*/k_BT)$ with an external potential $V_*= -  k_BT \ln (Z\, \rho _*)$, whose physical
origin is based on a  crude phenomenology, c.f.  \cite{sokolov,geisel}.
 With these re-definitions, the above transport  equation  takes the form  (21).

\section{Reverse engineering for Cauchy flights}

By choosing  $\mu  =1$  in the above,  we narrow  down  the whole discussion  to Cauchy processes, when e.g. (all integrals are evaluated by means of their Cauchy principal value)
 $$
(|\Delta |^{\mu /2} f)(x)\, =\, - \int_R [f(x+y) - f(x) - {{y\, \nabla f(x)}
\over {1+y^2}}]\, \nu _{\mu }(dy)
$$
$$\Downarrow $$
\begin{equation}
 (|\Delta |^{\mu /2} f)(x)\, =\, - \int  [f(x+y) - f(x) ] \nu _{\mu }(dy) \, .
\end{equation}
The Cauchy-L\'{e}vy measure,   associated with the  Cauchy  semigroup generator $|\Delta |^{1/2}\doteq |\nabla |$, reads
\begin{equation}
\nu  _{1/2}(dy) = {\frac{1}{\pi }} {\frac{dy}{y^2}}\, .
\end{equation}

By changing  an integration variable $y\rightarrow z=x+y$, we give  Eq. (20)  the familiar form
\begin{equation}
 (|\nabla | f)(x)\, =\, -  {\frac{1}{\pi }} \int  {\frac{f(z)- f(x)}{|z-x|^2}}\,  dz
\end{equation}
where $1/\pi |z-x|^2$ has an interpretation of an intensity with which jumps of the size $|z-x|$ occur.

\subsection{Ornstein-Uhlenbeck process}

In case of the Ornstein-Uhlenbeck-Cauchy (OUC) process, the drift is given by $b(x)= - \gamma x$, and an asymptotic
invariant density associated with
\begin{equation}
\partial _t \rho = - \lambda |\nabla | \rho + \nabla [(\gamma x)\rho ]
\end{equation}
 reads:
\begin{equation}
\rho _*(x) = {\frac{\sigma }{\pi (\sigma ^2 + x^2)}}
\end{equation}
where $\sigma = \lambda /\gamma $, c.f. Eq. (9) in Ref. \cite{olk1}

 A characteristic function of this density
reads $-F(p) = - \sigma  |p|$ and gives account of a    non-thermal fluctuation-dissipation balance.
The modified noise intensity parameter $\sigma $ is  a ratio of
 an intensity parameter  $\lambda $  of the  free
  Cauchy noise and  of the friction coefficient  $\gamma $.

From the start we know what is  the  drift $b(x)= - \lambda x$ which directs the process towards a target (stationary) pdf.
To deduce the Feynman-Kac potential ${\cal{V}}$ for the OUC   process,
 we need to evaluate
 \begin{equation}
{\cal{V}}(x)  = {\frac{\lambda }{\pi }} (\sigma ^2 + x^2)^{1/2}  \int \left[ {\frac{1}{\sqrt{\sigma ^2 + (x+y)^2}}} -
{\frac{1}{\sqrt{\sigma ^2 + x^2}}} \right] {\frac{dy}{y^2}} \, .
 \end{equation}

In the  notation $a=\sigma ^2 + x^2$,  $b= 2x$, $R(y)= \sigma ^2 +  (x+ y)^2$   indefinite integral  reads, \cite{gradstein}:
\begin{equation}
{\frac{\lambda }{\pi }}\,   \int\left[
 {\frac{\sqrt{a}}{y^2\sqrt{R(y)}}}  -  {\frac{1}{y^2}} \right] dy = {\frac{\lambda }{\pi }} \left[ -   {\frac{\sqrt{R(y)}}{y \sqrt{a}}}
 + {\frac{b}{2a}}\,  Arsh \left(
 {\frac{2a+by}{2\sigma |y|}}\right)
 + {\frac{1}{y}}\right]\, .
\end{equation}
Because of the singularity at $y=0$, we  must handle  the integral in terms of its principal value, i.e. by resorting to
$\int \rightarrow \int_{-\infty }^{-\epsilon } + \int_{\epsilon }^{+\infty }$, and  next  performing the $\epsilon \rightarrow 0$ limit.

Taking into account that $arsinh \, x \equiv  \ln (x+ \sqrt{1+x^2})$, \cite{stephanovich},  we ultimately  get
\begin{equation}
{\cal{V}}(x) = {\frac{\lambda }{\pi }} \left[ - {\frac{2}{\sqrt{a}}} + {\frac{x}{a}}\ln {\frac{\sqrt{a}+x}{\sqrt{a}-x}}\right] \, .
\end{equation}
${\cal{V}}(x)$  is bounded both from below and above, with the asymptotics  $(2/|x|) \ln |x|$ at infinities, well fitting to the
 general mathematical construction  of (topological) Cauchy processes in external potentials, \cite{olk}.

  Accordingly, we  know for sure that there exists  a jump-type
  process  driven by  the Cauchy semigroup  with the potential function  ${\cal{V}}$, Eq. (29), whose invariant density coincides with that
 for the Langevin-supported  OUC process.  This form of the semigroup potential, gives a guarantee that $\rho _*$ actually  \it is \rm an
 asymptotic invariant density of the  process. In Fig. 1 we reproduce the functional shape of the potential (29), \cite{stephanovich}.
 \begin{figure}
\begin{center}
\includegraphics [width=0.6\textwidth]{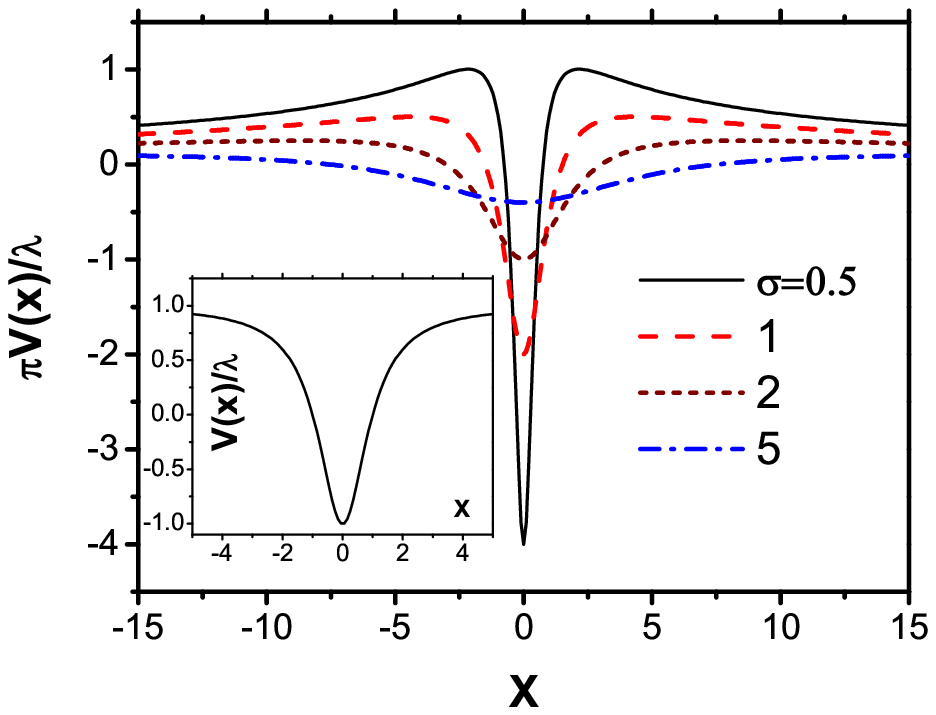}
\includegraphics [width=0.6\textwidth]{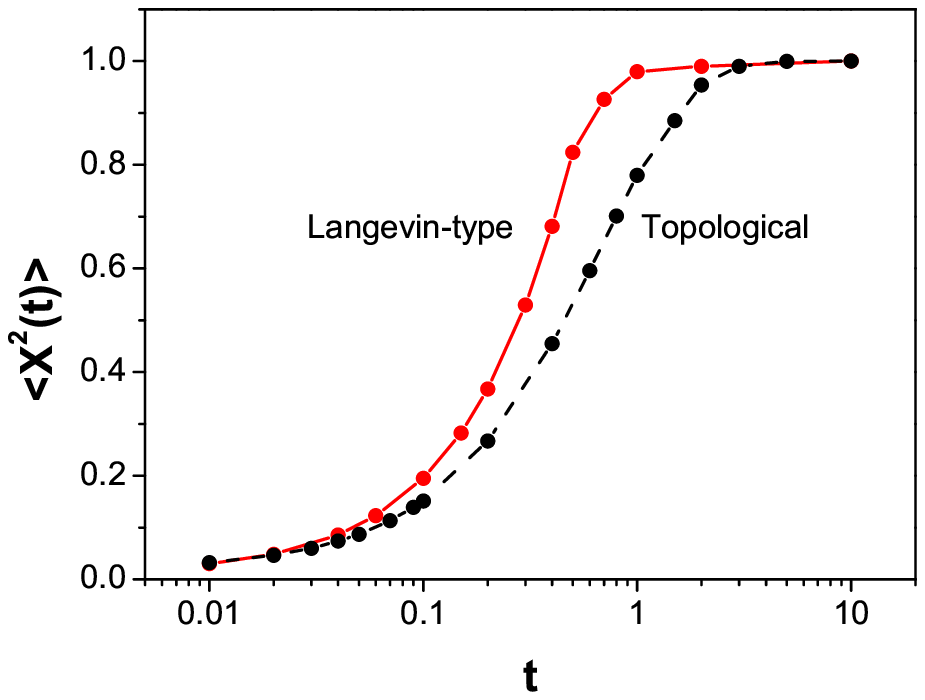}
\end{center}
\caption{Upper panel: the coordinate dependence of semigroup  potentials ${\cal{V}}(x)$:
(29) for different $\sigma$  and (33)
(inset). Lower panel:  time dependent variance $X^2(t)$
for Langevin-type (solid line) and semigroup-driven  (dashed line)
processes associated with the pre-defined  target pdf (33). Points correspond to numerical calculation.}
\end{figure}

 \subsection{Confined Cauchy process}

 The  OUC process  is not  confined, since for  the Cauchy  density its second moment is nonexistent.
We shall  adopt  the OUC discussion  to  Cauchy-type processes whose invariant densities admit  the second  moment.
Let us   consider the quadratic Cauchy density:
\begin{equation}
\rho _*(x) = {\frac{2}{\pi }}\, {\frac{1}{(1+x^2)^2}}\, .
\end{equation}
The action of $|\nabla |$ upon this density can be  evaluated
by recourse to the free Cauchy  evolution.

We note  that  $  (1/\sqrt{2\pi })\rho_*^{1/2} =
(1/\pi )/(1+x^2)$  actually is the Cauchy probability  density.
Let us  consider $f(x) = \rho _*^{1/2} $ as the initial data for
the free Cauchy evolution $\partial _t f = \lambda  |\nabla |f$.
 This   takes $f(x)$ into
\begin{equation}
f(x,t)= {\frac{2}{\pi }} \, {\frac{1 + \lambda t}{[(1 + \lambda t)^2 + x^2]}} \, .
\end{equation}
Since
\begin{equation}
 \lambda |\nabla |f =  - \lim_{t\downarrow 0} \partial _t f
\end{equation}
we end up with
\begin{equation}
{\cal{V}} (x) = {\frac{\lim_{t\downarrow 0} \partial _t f}{f}}(x) = \lambda  {\frac{x^2-1}{x^2 +1 }}\, .
\end{equation}
A minimum $ -\lambda $ is achieved at $x=0$, ${\cal{V}}=0$  occurs for $x=\pm 1$,
a maximum $+\lambda $ is reached at $x \rightarrow  \pm \infty $. The functional shape of the potential is depicted as an inset in Fig. 1.

The potential is bounded both from below and above, hence can trivially be made non-negative (add $\lambda $).
 Therefore,  the invariant  density  (30) is  fully  compatible with the general construction of
 the Cauchy-Schr\"{o}dinger semigroup and the induced jump-type process,  c.f. Corollary 2, pp. 1071 in \cite{olk}.
  This topological Cauchy  process is
  induced by the Cauchy generator plus a potential function ${\cal{V}}$ given by Eq. (49).
   c.f. Corollary 2, pp. 1071 in \cite{olk}.
 The process is of the  jump-type  and  can be obtained as an $\epsilon \downarrow 0$ limit of a step process, e.g. jump process
 whose jump size is bounded from below  by $\epsilon >0$ but unbounded from above.

In connection with the  reverse engineering problem of Ref. \cite{eliazar} let us note that if  the  quadratic Cauchy  density   actually stands for
  a stationary density   of  the  fractional Fokker-Planck equation  with a drift, then  we should  have:
  \begin{equation}
\partial _t \rho _* = 0 = - \nabla (b\, \rho _*) - \gamma |\nabla |\rho _* \, .
  \end{equation}
Therefore the drift function  may be  deduced by means of an indefinite integral:
\begin{equation}
b(x) = -  {\frac{\gamma }{\rho _*(x)}} \int (|\nabla |\rho _*)(x)\, dx = -{\frac{\gamma }8}\,  (x^3 +  3 x)  \, .
\end{equation}

For clarity of discussion, in the lower panel of Fig. 1,  we report a comparison of dynamical patterns of behavior for the semigroup-driven and Langevin-induced  scenarios
beginning from common (delta-type) initial data and approaching a common (pre-defined) target  pdf (33).

{\bf Remark:} It is worthwhile to notice that   potential functions  for the drift $b= - \nabla U$ of the form $U(x)=a x^2 + bx^4$ were investigated in Ref. \cite{klafter} with a focus on bimodality of the resultant stationary pdfs.  For example a quartic potential $U= bx^4$ is known to induce $\rho _*(x)= [\pi (1-x^2+x^4)]^{-1}$.
Our unimodal density (33) belongs to the subclass associated with the just mentioned $U(x)$.

\subsection{Confined Cauchy family}

We may  consider  various  probability densities  as  trial ones. Let us  pay attention to a broader class of densities
 that bear  close  affinity  with the   Cauchy noise.
With a given continuous probability distribution $\rho $  we associate  its Shannon entropy
 $S(\rho ) = -\int \rho \, \ln \rho \, dx$, \cite{kapur}.
  If an expectation value $E[\ln (1+x^2)]$ is prescribed (e.g. fixed),
  the maximum entropy probability function belongs to a one-parameter family
 \begin{equation}
\rho _* (x)= {\frac{\Gamma (\alpha )}{\sqrt{\pi } \Gamma (\alpha -1/2))}}\, {\frac{1}{(1+x^2)^{\alpha }}}
 \end{equation}
where $\alpha  >1/2$, \cite{kapur}.

 The gamma function $\Gamma (\alpha ) = \int_0^{\infty } \exp(-t)\, t^{\alpha -1}\, dt $ we specialize to integer
$\alpha = n+1$-values, with $n\geq 0$. Then $\Gamma (n+1) = n!$  and $\Gamma (\alpha -1/2) \rightarrow
\Gamma (n+ 1/2)= [(2n)!\sqrt{\pi }]/ n!2^{2n}$.

As an exemplary case let us consider
\begin{equation}
\rho _* (x) = {\frac{16}{5\pi }} \, {\frac{1}{(1+x^2)^4}} \,
\end{equation}
By adopting the  previous procedure,  c.f. \cite{gradstein}, and evaluating the principal value integrals,  we end up with the following expression for the  Cauchy semigroup  potential:
\begin{equation}
{\cal{V}}(x)= {\frac{\gamma }{2(1+x^2)}}\,  (x^4 +6x^2 -3)\, .
\end{equation}

The potential is bounded from below, its minimum at $x=0$ equals $-3\gamma /2$.  For large values of $|x|$, the potential behaves as
$\sim (\gamma /2) x^2$ i.e. shows up a harmonic behavior.

Apart from  the unbounded-ness  of ${\cal{V}}(x)$ from above, this potential
obeys the minimal requirements of Corollary 2 in Ref. \cite{olk}: can be made positive (add a suitable constant),
 is  locally bounded (e.g. is bounded  on  each compact set) and is measurable (e.g.   can be arbitrarily well
  approximated by means of sequences of step functions).
The  Cauchy generator plus the  potential (38)   determine uniquely an associated Markov process of the jump-type and its step process approximations.

We can readily address the reverse engineering problem of Ref. \cite{eliazar}. For the density (37), we ultimately  get,
\cite{stephanovich}:
\begin{equation}
b(x) = - {\frac{\gamma x}{16}}\, (5x^6 + 21x^4 + 35 x^2 +35)\, .
\end{equation}
This a bit discouraging  expression  which  shows  a linear friction  $b\sim - x$  for small $x$ and a strong taming behavior
 $b\sim -x^7$ for large $x$,  still fits to  the above mentioned Corollary 2 of Ref. \cite{olk}.

\section{Summary}

We have  generalized the reverse engineering (targeted stochasticity) problem of Ref. \cite{eliazar} beyond the
original L\'{e}vy-Langevin  processes setting. We have demonstrated that the notion of L\'{e}vy flights in confining potentials
is not limited to the Langevin scenario.
 The L\'{e}vy-Schr\"{o}dinger semigroup involves  the notion of external potentials as well. But  then with no link to
  any standard Langevin representation.

  Our version of the reverse engineering problem amounts to reconstructing from a given (target) stationary density
  the potential functions that either: (i) define the forward drift of the  Langevin process, or (ii) enter the
  Schr\"{o}dinger-type Hamiltonian expression  in the  semigroup dynamics. Both dynamical scenarios are expected to yield
  the same asymptotic outcome i.e. the  pre-selected target pdf. This goal can be achieved in the semigroup  picture (and models of
   an impact of  inhomogeneous environments upon  L\'{e}vy flights) only if suitable restrictions on the semigroup potentials are observed.
 The relevant mathemtical hints come from Ref. \cite{olk} and were illustrated for the case of Cauchy driver.

 Insightful, explicitly  solvable  models  are scarce in   theoretical studies of L\'{e}vy flights, in
 the presence of external potentials  and/or  external conservative  forces. Therefore, our major task was
 to find novel analytically tractable    examples,  that would shed some  light on apparent discrepancies between dynamical patterns of
  behavior   associated   with   two  different   fractional  transport   equations (15) and (20) that  are met in the literature
   on L\'{e}vy flights. Albeit the predominant part of this research is devoted to the standard Langevin modeling.

 We note that  a departure point for our investigation was a familiar transformation of the Fokker-Planck operator into
  its Hermitian (Schr\"{o}dinger-type) partner, undoubtedly valid in the Gaussian setting.
  The Fokker-Planck and the corresponding parabolic equation (plus a compatibility condition (5)) essentially describe the same  random dynamics.

 An analogous transformation is non-existent for non-Gaussian processes.
  The two fractional Fokker-Planck equations (13) and (15),  are inequivalent in the non-Gaussian setting,
   hence the semigroup dynamics and the  Langevin dynamics with the L\'{e}vy driver (e.g. noise) refer to different random processes.
This behavior we have depicted in the lower panel of Fig. 1. The main technical reason of the incongruence of the two processes seems to be rooted in that the stable noise generator is a (non-locally defined)  pseudo-differential operator, while the standard Laplacian
(Wiener noise generator) is locally defined.
The reverse engineering problem allowed us to demonstrate that  those two processes may nevertheless share the same target pdf and may interpolate
 between common pairs of boundary (initial and terminal) pdfs.  Albeit in a different dynamical fashion.

One may wonder whether there is some symmetry principle (like e.g. a  remnant of the time-symmetric formulation of the Schr\'{o}dinger boundary data problem, \cite{klauder,olk,zambrini,zambrini1} that allows to relate two fixed  boundary densities by means of different dynamical scenarios. Actually, our observation that the semigroup-driven and Langevin-driven jump-type processes may share a common invariant pdf,  that in turn  is dynamically  accessible from a commmon for both  processes initial pdf, stands for an indirect proof that the involved dynamical scenarios are different. The assumption about driving mechanisms is the only freedom left in the above mentioned boundary data problem, once the initial and terminal pdf data are chosen.

\end{document}